\documentclass[prb,aps,twocolumn,amsmath,amssymb,showpacs]{revtex4}

\usepackage{dcolumn}
\usepackage{bm}
\usepackage{graphicx}
\usepackage{psfrag}
\usepackage{subfigure}

\begin{document}
 \title{Fractional charges and spin-charge separation in one-dimensional Wigner lattices}
\author {M. Daghofer and P. Horsch}
 \affiliation{ Max-Planck-Institut f\"ur Festk\"orperforschung,
               Heisenbergstrasse 1, D-70569 Stuttgart, Germany }
 \date{March 26, 2007}

\begin{abstract}
We study density response $N(k,\omega)$ and one-particle
spectra $A(k,\omega)$ for a Wigner lattice model at quarter filling using
exact diagonalization. We investigate these observables for models with short and long-range
electron-electron interaction and show that truncation of the electron
repulsion can lead to very different results. The spectra show clear signatures of charge fractionalization
into pairs of domain walls, whose interaction can be attractive or repulsive
and is controlled by the formal fractional charges. In striking contrast to
a bound exciton in $N(k,\omega)$, we find an antibound quasi-particle in
$A(k,\omega)$, which undergoes spin-charge separation. 
We present a case of extreme particle-hole asymmetry, where
photoemission shows spin-charge separation, while inverse photoemission
exhibits an uncorrelated one-particle band.\\
$[$ published in Phys. Rev. B {\bf 75}, 125116 (2007) $]$
\end{abstract}


\maketitle

\section{Introduction}
\label{sec:intro}

In contrast to higher dimensions, where interacting electrons are renormalized
into `Landau quasi-particles' that near the Fermi surface still `look' like 
electrons, 
drastic things can happen in one dimension.~\cite{Maekawa01}
The elementary excitations are collective ones involving many electrons, \emph{spin and
charge separate} and move independently of each other, as seen, e.g., in
photoemission (PES) experiments.\cite{Kim06} Another case where 
many-body effects lead to an apparent splitting of elementary particles
is \emph{quantum number fractionalization}.\cite{Oshikawa06} Probably its most
famous realization is the fractional quantum Hall
effect.\cite{Lau83} Likewise, Peierls distorted~\cite{Heeger88} and
charge-ordered~\cite{Hubbard78,Rice82} one-dimensional (1D) systems
with degenerate ground states have    
fractionally charged solitons or domain walls as elementary excitations. 
In fact, quantum number fractionalization
in one and two dimensions are intimately related.\cite{Seidel06}
This Paper presents a study of dynamic observables 
for a model showing both effects,
spin-charge separation as well as quantum number fractionalization. 

When (long-range) Coulomb interaction is the dominant
energy scale of a system,  electrons try to minimize their energy by
maximizing their distance and crystallize into a
Wigner lattice (WL).\cite{Wigner34} Hubbard suggested this mechanism in
the context 
of TCNQ charge-transfer salts~\cite{Hubbard78} where it is, however, difficult
to distinguish from a $4k_F$ charge-density wave driven by a Fermi surface
instability.~\cite{Hubbard78,Rice82,Nad06}
As pointed out recently,~\cite{Horsch05}  
longer-range hopping changes the Fermi-surface topology in doped edge-sharing CuO-chain 
compounds like Na$_{1+x}$CuO$_2$~\cite{Sofin05} and
Ca$_{2+y}$Y$_{2-y}$Cu$_5$O$_{10}$,~\cite{Kudo05} and this allows for a clear
distinction between the Fermi-surface independent WL and the charge-density wave. 

The elementary excitations of a WL  consist of domain walls (DWs).
Their fractional charge follows from topological 
arguments~\cite{Jackiw75,Hubbard78,Rice82} and merely 
reflects the $n$-fold ground-state degeneracy at
$x=m/n$ filling -- it is not related to the specific form of the  
interaction  that generates the charge order.
After introducing the employed model Hamiltonian in Sec.~\ref{sec:model}, 
we show in Sec.~\ref{sec:spinless} that these fractional charges have direct physical meaning
in a model with long-range Coulomb repulsion among electrons, because the
resulting effective interaction between DWs is also Coulomb-like with a coupling
constant determined by the fractional charges, their signs and
distance. Consistent with the formal charges, the interaction is
\emph{attractive} for the charge response~\cite{Valenzuela03,Fratini04,Mayr06}  
but \emph{repulsive} for the electron addition (removal) process. 
We present and discuss charge response $N(q,\omega)$ and spectral densities
$A(k,\omega)$ and show that the correspondence of formal DW charge and
effective physical charge is tied to the long-range nature of
Coulomb repulsion and absent for models with truncated electron-electron interaction.

As the energy scale for spin excitations is much smaller
than either the Coulomb repulsion or the kinetic energy, previous
investigations considered WL formation of spinless fermions. 
We will address the spin degree of freedom in Sec.~\ref{sec:spin} and we will show how the interplay of spin and
DW excitations leads to two different scenarios for the spectral
density: (i) For nearest neighbor hopping, the coherent antibound DW
excitation undergoes spin-charge separation in perfect analogy to an electron
or hole added to the half-filled 1D Hubbard model. (ii) For 2nd 
neighbor hopping, we find  striking differences between the particle and hole
channels, where an electron behaves like an independent particle while a hole
decays into a spinon and a holon.

\section{Model and methods}
\label{sec:model}

We study the Hubbard-Wigner Hamiltonian~\cite{Mayr06}  
\begin{equation}\begin{split}\label{eq:hamiltonian}
H & = t_1 \sum_{i,\sigma} ({c}^\dagger_{i, \sigma}{c}^{\phantom{\dagger}}_{i+1, \sigma} + \textrm{h.c.})+
    t_2 \sum_{i,\sigma} ({c}^\dagger_{i, \sigma}{c}^{\phantom{\dagger}}_{i+2, \sigma}  + \textrm{h.c.}) \\
  &\quad +  U \sum_i n_{i, \uparrow}n_{i, \downarrow}
   + \sum_{l = 1}^{l_\textrm{max}} V_l \sum_i (n_i- \bar n)(n_{i+l}-\bar n)\;,
\end{split}\end{equation} 
with nearest neighbor (NN) and next nearest neighbor (NNN) hopping 
amplitudes $t_1$ and $t_2$. 
Operator ${c}^\dagger_{i, \sigma}$ (${c}^{\phantom{\dagger}}_{i,\sigma}$) creates (destroys) an electron
with spin $\sigma$ at site $i$, $n^{\phantom{\dagger}}_{i, \sigma} = {c}^\dagger_{i,
  \sigma}{c}^{\phantom{\dagger}}_{i, \sigma}$ and $n_i = n_{i, \uparrow}+
n_{i, \downarrow}$ give the density, $\bar n$ is the average density. 
We focus on long-range Coulomb repulsion $V_l=V/l$~\cite{note2}, but also
discuss truncated interaction ($l_\textrm{max} = 3$) to illustrate that
truncation may have a strong impact on the excitation spectra.
The NN Coulomb matrix element $V$ is used as unit of energy.
We treat chains up to $L= 24$ in the spinless case and $L=16$ with spin using
Lanczos diagonalization.

To our knowledge, neither the single-particle spectral function $A(k,\omega)$
nor the density response $N(q,\omega)$ of a Wigner lattice have been studied
before. We consequently treat here the most transparent case given by 
quarter filling, i.e., one electron per two sites.
We will also analyze an effective low-energy model in terms of domain walls
that is appropriate for small hoppings $t_1, t_2\ll
V,U$,~\cite{Mayr06,Valenzuela03,Fratini04} and compare the results to those of
the full model (\ref{eq:hamiltonian}) that is formulated in  terms of
electrons or spinless fermions. 

\section{Dynamics of the spinless model}
\label{sec:spinless}

\begin{figure}
  \centering
  \psfrag{a}{\hspace*{-1em}(a)}
  \psfrag{b}{\hspace*{-1em}(b)}
  \psfrag{c}{\hspace*{-1em}(c)}
  \psfrag{t}{$t_1$}
  \includegraphics[width=0.4\textwidth]{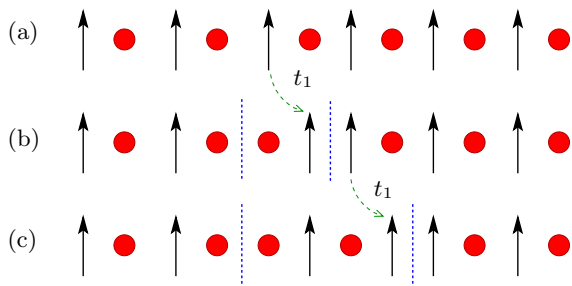}\\
\caption{(Color online) One of the two ground states (a) with circles denoting
  holes and solid arrows spinless fermions, pair of domain walls (indicated by
  dashed vertical lines) at distance
  $d=2$ created by moving one electron via $t_1$ (b), configurations with
  distance $d=4$ between the DWs (c). \label{fig:dws_pm}}
\end{figure}

At quarter-filling 
the ground state of (\ref{eq:hamiltonian}) is twofold degenerate for spinless
fermions, and the lowest
excitations are therefore given by domain walls with a formal charge $\pm
e/2$.~\cite{Jackiw75,Hubbard78,Rice82}  The existence of such DWs as 
elementary excitations follows from topological considerations and only depends on
the degeneracy of the ground state, not on the details of the
Hamiltonian or the interaction that stabilizes the ground state. In the WL, where the hopping is not large
enough to destroy the charge order (the gap is expected to vanish at $t_1/V
\sim 0.2$~\cite{Capponi00}), the description in terms of domain walls is
useful, and we will show that the dynamics of the WL can be described by
DWs and their interaction. DWs can be created from the perfectly ordered state
by a NN hopping process $t_1$, see Fig.~\ref{fig:dws_pm} (a) and (b). Creating
a pair of domain walls is penalized by $V$ and costs potential energy. DWs can
move by  $t_1$, see Fig.~\ref{fig:dws_pm} (b) and (c), and their potential
energy can depend on the distance $d$ between the two DWs. In
Refs.~\onlinecite{Valenzuela03,Fratini04}, the DW interaction for a model with
long-range Coulomb repulsion has been discussed and been found to correspond
to a Coulomb-like attraction as it would be expected between two physical
charges of $\pm e/2$ each.  In Ref.~\onlinecite{Mayr06}, an effective
low-energy Hamiltonian in terms of DWs has been analyzed.

\begin{figure}
  \centering
  \psfrag{Nqo}{\hspace*{-1em}$N(q,\omega)$}
  \includegraphics[width=0.4\textwidth]{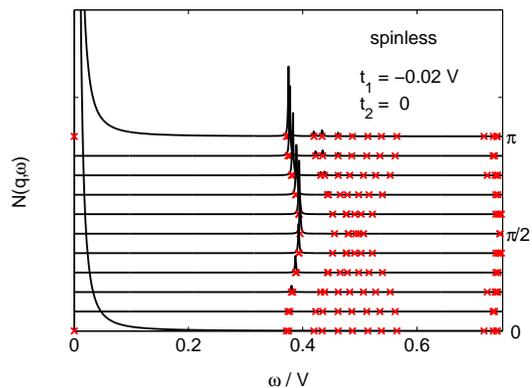}
  \caption{(Color online) Density response $N( q,\omega)$ with $t_1 = -0.02 V$
  and $t_2 = 0$ for different momenta $q=\pi m/10,\ m = 0,\dots, 10$. The $\times$
  show the eigen energies relative to the ground-state energy.\label{fig:rho_spinless}} 
\end{figure}

The lowest eigen energies of (\ref{eq:hamiltonian}) with a small NN
hopping $t_1 = -0.02$ can be seen in Fig.~\ref{fig:rho_spinless}
($\times$). At $\omega =0$ and momenta $k=0$ and $k=\pi$, we see the two
ground states. While this degeneracy at finite $t_1$ is only perfect
in the thermodynamic limit, the numerical data in Fig for a $N=20$ site
ring shows that the states at    $k=0$ and $k=\pi$ are almost degenerate.
The collection of states at $\omega \sim 0.5 \pm 0.08$ have been shown to
represent a continuum of two independent domain walls,~\cite{Mayr06} 
and the states above
 $\omega\sim 0.8$ mark the beginning of the four-DW continuum. In addition
to the continua, we find an exciton corresponding to a bound 2-DW state at
energies just below the 2-DW continuum.~\cite{Mayr06}

\begin{figure}
  \centering
  \psfrag{Nqo}{\hspace*{-1em}$N(q,\omega)$}
  \includegraphics[width=0.4\textwidth]{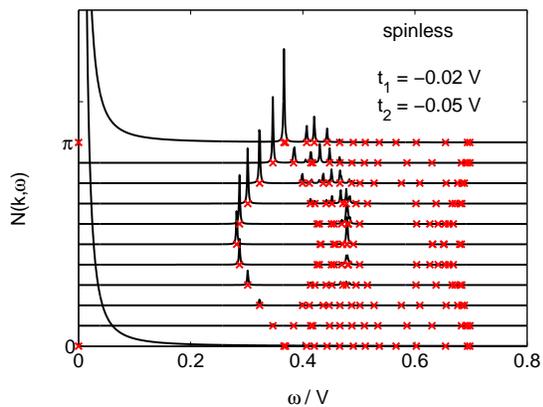}
  \caption{(Color online) Density response $N( q,\omega)$ with $t_1 = -0.02 V$
  and $t_2 = -0.05$ for different momenta $q=\pi m/10,\ m = 0,\dots, 10$. The $\times$
  show the eigen energies relative to the ground-state energy.\label{fig:rho_spinless_t1_t2}} 
\end{figure}

In addition to the eigen energies, Fig.~\ref{fig:rho_spinless} shows the
dynamic charge structure factor   
\begin{equation}
N( q,\omega) =  \sum_{m} 
|\langle m|\rho_{ q} | \phi_0\rangle|^2
\delta(\omega - (E_m -E_0))\;,
\end{equation}
where $|m\rangle$ and $E_m$ are the eigenstates and energies of the Hamiltonian, and
$|\phi_0\rangle$ is the ground state with energy $E_0$.  
For the perfect WL without quantum fluctuations, it shows only signals at $\omega = 0$
and momenta $ k = \pi$ and $ k = 0$. Spectral weight at finite-frequency is
produced by fluctuations around perfect charge order and Fig.~\ref{fig:rho_spinless}
reveals that most spectral weight is observed in the exciton, while the two-DW
continuum at $0.4 V\lesssim\omega\lesssim0.6 V$ contains almost no weight. 

In edge-sharing Cu-O chain compounds like the quarter-filled Na$_3$Cu$_2$O$_4$
system,  the NNN hopping $t_2$ is expected to be  larger than $t_1$,~\cite{Horsch05}. 
The eigenvalues and charge response for $t_1 = -0.02 V, t_2 =
-0.05 V$ are shown in Fig.~\ref{fig:rho_spinless_t1_t2} for a 20-site ring. 
As explained in
Ref.~\onlinecite{Mayr06}, the bound exciton has in this case its minimum at
$q = \pi/2$. 
At a critical value of $t_2$ the exciton gets soft and
the WL with periodicity $\pi$ is destroyed and charge order with a
periodicity $\pi/2$ sets in. Again, we see that the exciton has large
weight in $N(q,\omega)$ and should in principle be observable in charge response.

\begin{figure}
  \centering
  \vspace*{-0.5em}
  \subfigure{\includegraphics[width=0.4\textwidth]
    {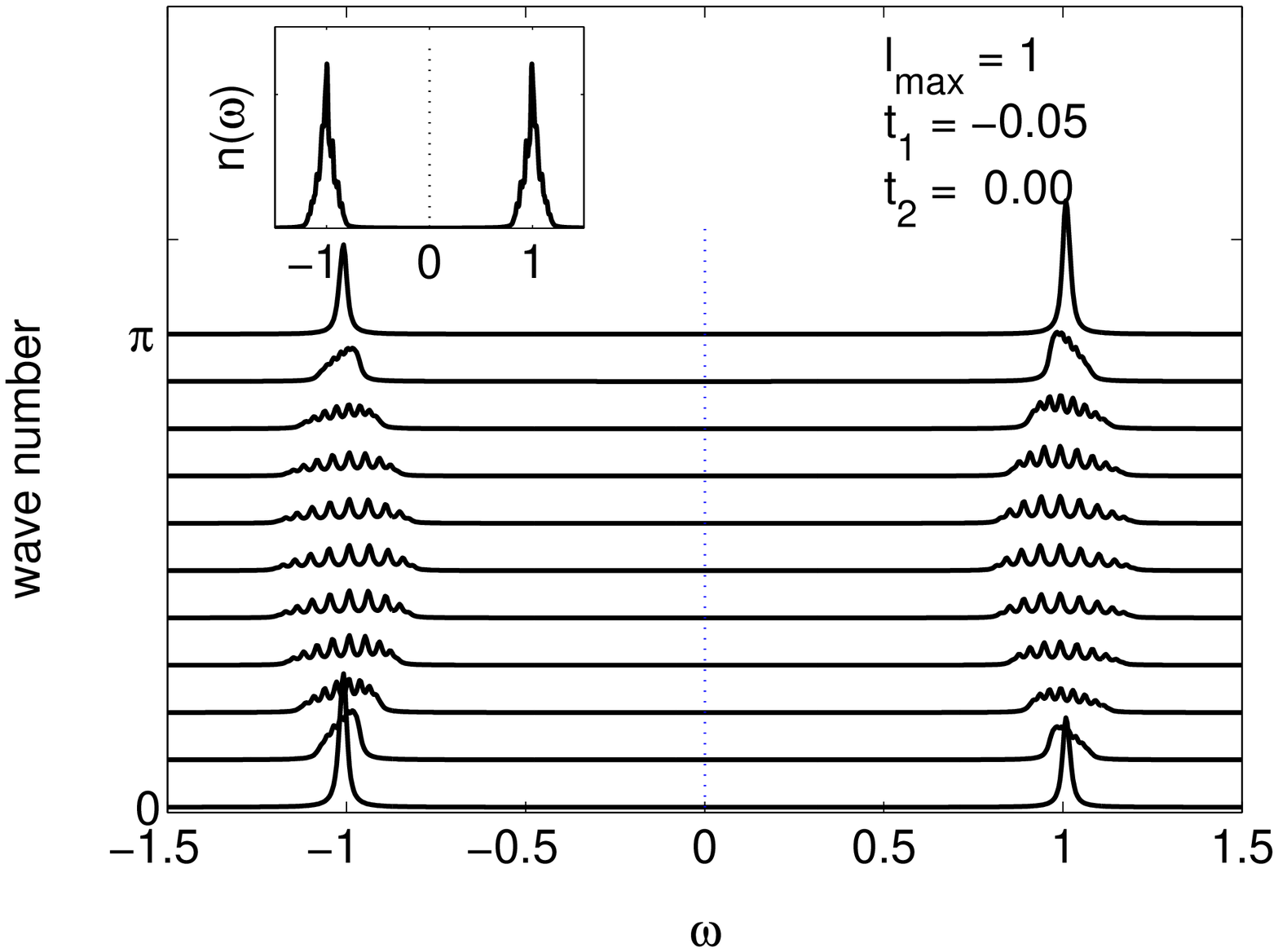}\label{fig:spec_spinless_lmax1}}\\[-4em]
  \subfigure{\includegraphics[width=0.4\textwidth]
    {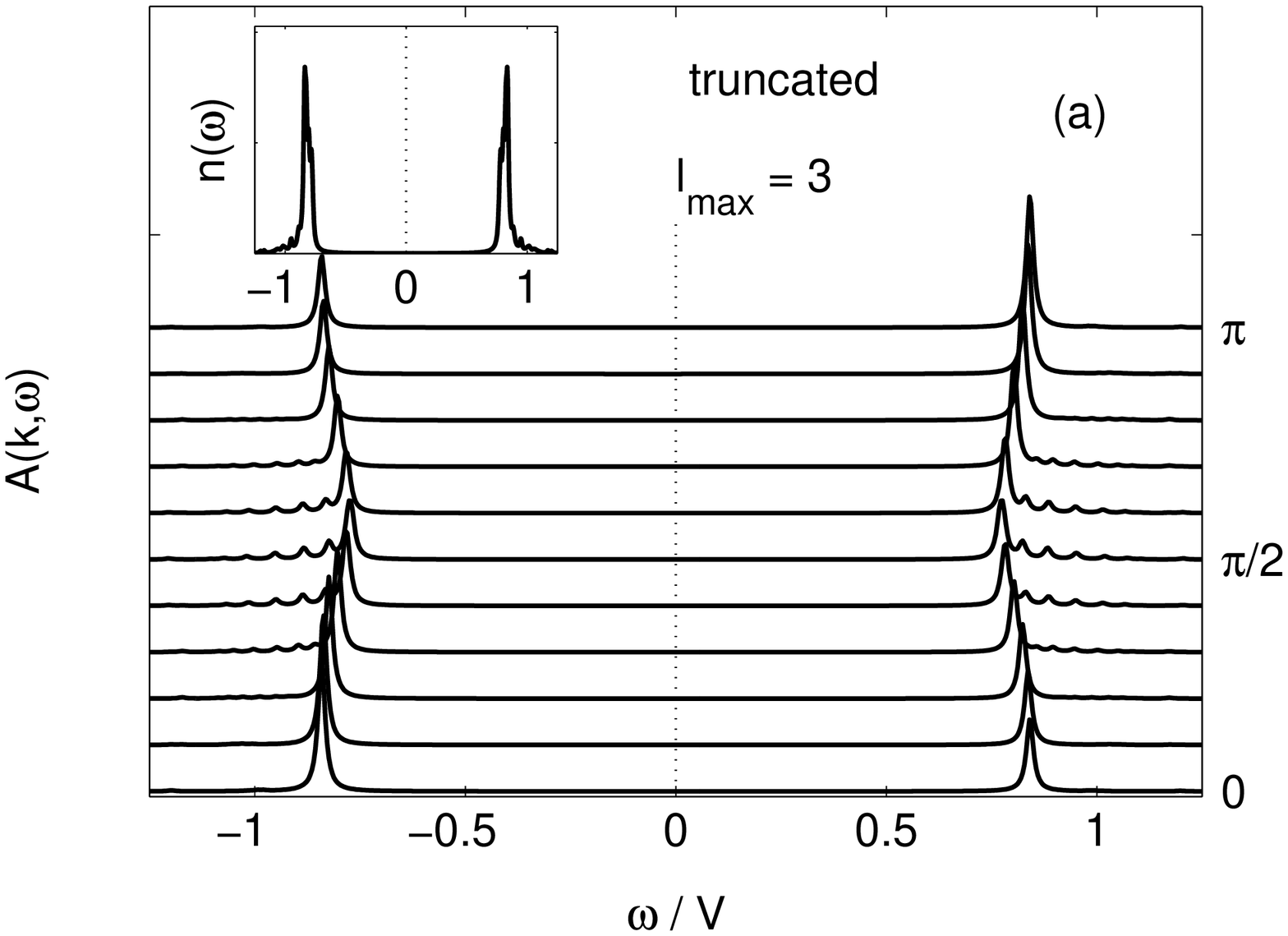}\label{fig:spec_spinless_lmax3}}\\[-4em]
  \subfigure{\includegraphics[width=0.4\textwidth]
    {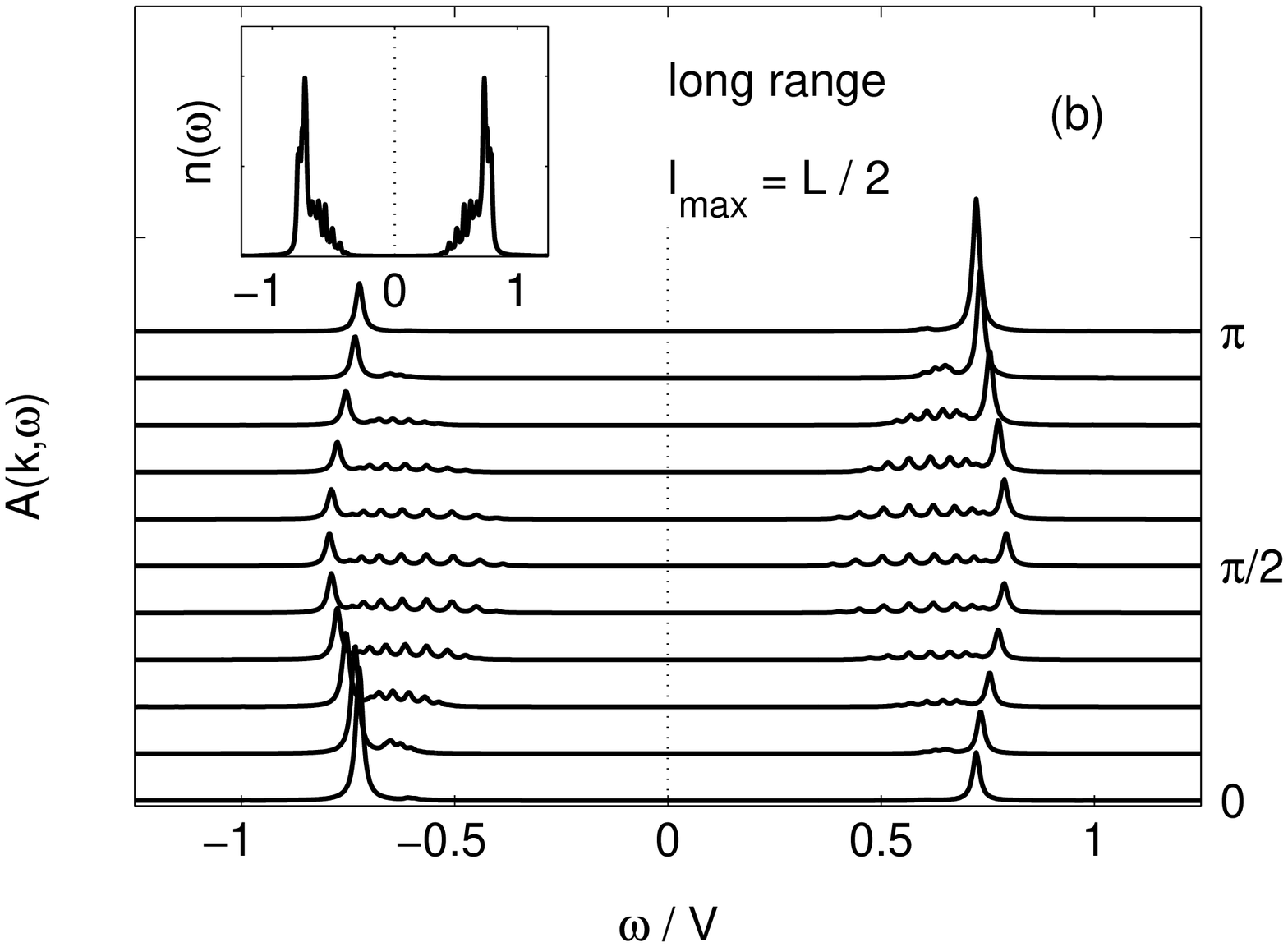}\label{fig:spec_spinless_coulomb}}\\[-1.5em]
  \caption{Spectral density $A(k,\omega)$ for 
  spinless fermions for (a) truncated $V_l$  with $l_{max}=1$, (b) truncated
  $V_l$  with $l_{max}=13$ and (c) long-range $V_l$ with $t_1 = -0.05 V, t_2 
  = 0$. Insets show the density of states. \label{fig:spec_spinless}} 
\end{figure}

Interestingly, the relevant lowest excitations of the WL continue to be  pairs
of DWs in the case of the one-particle spectral density
\begin{equation}\begin{split}\label{eq:ak}
A( k, \omega)&  =  \sum_{m} 
|\langle m^+| {c}^\dagger_{ k, \uparrow} | \phi_0\rangle|^2
\delta(\omega - (E^+_m -E_0))\\ 
&\quad +  \sum_{m} 
|\langle m^-| {c}^{\phantom{\dagger}}_{ k, \uparrow} | \phi_0\rangle|^2
\delta(\omega - (E^-_m -E_0))\;,
\end{split}\end{equation}
where $|m^+\rangle$  ($|m^-\rangle$) are eigenstates with eigenenergies
$E^+_m$ ($E^-_m$) of the Hamiltonian with one particle added (removed). This
observable is shown in Fig.~\ref{fig:spec_spinless} for a small NN hoping $t_1
= -0.05V$  and three different cases
for the interaction in (\ref{eq:hamiltonian}): (i) NN repulsion
only ($l_\textrm{max}=1$), where we see only a broad featureless continuum,
(ii) slightly  longer-range interaction with $l_\textrm{max} = 3$, where we
find a sharp quasiparticle below a continuum with small weight, and (iii)
long-range Coulomb-repulsion with $l_\textrm{max} = L/2$, where the quasiparticle is
observed \emph{above} the continuum. We will now go on to show how these
spectra - the continuum, the bound quasiparticle and the antibound
quasiparticle - 
which have been obtained without any further assumption
by exact diagonalization of the Hubbard-Wigner model,
can be understood in terms of DWs and their interaction.

\begin{figure}
  \centering
  \psfrag{a}{\hspace*{-1em}(a)}
  \psfrag{b}{\hspace*{-1em}(b)}
  \psfrag{c}{\hspace*{-1em}(c)}
  \psfrag{t}{$t_1$}
  \includegraphics[width=0.4\textwidth]{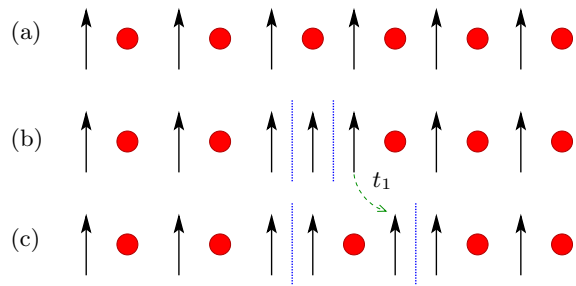}\\
\caption{(Color online) One of the two ground states (a) with circles denoting
  holes and solid arrows spinless fermions, pair of domain walls (indicated by
  dashed vertical lines) at distance
  $d=1$ created by \emph{adding one electron} (b), configurations with
  distance $d=3$ between the DWs mediated by $t_1$(c). \label{fig:dws_p}}
\end{figure}

\begin{figure}
  \centering
  \includegraphics[width=0.4\textwidth, viewport=10 0 600 420,clip]{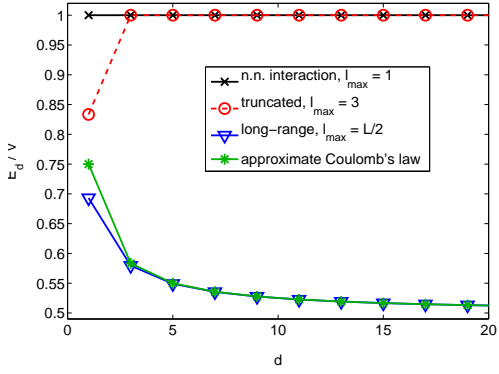}\\
  \caption{(Color online) Potential $E_d$ between two domain walls at distance
  $d$, see Fig.~\ref{fig:dws_p}, for
  long-range electron-electron interaction ($l_\textrm{max} = L/2$,
  $\triangledown$), effective DW interaction (\ref{eq:eff_DW_int}) 
  ($\ast$), and for truncated electron-electron interaction ($l_\textrm{max}
  = 1$, $\times$ and $l_\textrm{max}
  = 3$, $\circ$).\label{fig:pot_dw} }
\end{figure}

The schematic illustration in Fig.~\ref{fig:dws_p} shows that
an additional electron again creates two DWs, this time both with an identical
charge of $-e/2$. Again, their formal charge does not depend on details of the
Hamiltonian: The total system has charge $-e$ (the additional electron) and both DWs are
equivalent, which gives each DW a formal charge $-e/2$. 
The interaction $E_d$ between two domain walls at distance $d$ can be obtained
by calculating the potential energy for a configuration where they are $d$
sites apart (as shown in Fig.~\ref{fig:dws_p} for distances $d=1,3$) by
setting all hoppings to zero. The
resulting interaction is depicted in Fig.~\ref{fig:pot_dw} both for long-range
and for truncated electron-electron interaction. If the Hamiltonian
(\ref{eq:hamiltonian}) has only NN Coulomb repulsion $l_\textrm{max} =1$, the
energy ($\times$ in Fig.~\ref{fig:pot_dw}) does not depend on $d$, and the DWs
are therefore independent. For $l_\textrm{max} =3$, the DWs are independent
for large distance and \emph{attractive} for small $d$. This means that their
interaction has the \emph{opposite sign} from the one expected from their
formal charges of $-e/2$ each. 
For the long-range case $l_\textrm{max} =
L/2$ ($\triangledown$  in Fig.~\ref{fig:pot_dw}), however, the 
asymptotic behavior of the interaction between two DWs  
decays like their inverse distance $d$: 
\begin{equation}\label{eq:eff_DW_int}
E_d/V \sim 1/2 + e^2/(4d)\;.
\end{equation}
Hence, the interaction between the DWs is Coulomb-like and the prefactor
is given by the  two fractional charges of $-e/2$.
The asymptotic relation ($\ast$ in Fig.~\ref{fig:pot_dw}) gives an
excellent approximation already for rather small DW distances $d>1$. 
That (formal) fractional charges are the elementary excitations in charge
ordered chains can be derived from the ground-state degeneracy
alone,~\cite{Hubbard78} without any reference to the 
particular form of the interaction stabilizing the degenerate ground states.  
In the case of \emph{long-range} electron-electron interaction, and
\emph{only} in this case, the formal fractional charges have a very direct
physical meaning: Their interaction corresponds exactly to that expected for `half-electrons'.

For the description of the photoemission process of 
a spinless fermion added to the perfect WL an effective low-energy
Hamiltonian in terms of DWs can be obtained that contains hopping $t_1$ and
the potential energy $E_d$. A reader interested in details for the following
short derivation should consult Ref.~\onlinecite{Mayr06}, where an analogous
treatment is given for the case depicted in Fig.~\ref{fig:dws_pm}. 
Starting from the ground state with $p=0$ 
\begin{equation}\label{eq:ground_WL}
|\phi_\textrm{WL}\rangle = \frac{1}{\sqrt{2}} (|0 \uparrow 0 \uparrow 0
 \uparrow \dots \rangle + |\uparrow 0 \uparrow 0 \uparrow 0 \dots \rangle ) 
\end{equation}
and adding an electron with momentum
$k$, we arrive at a state $|\psi_{d=1,k}\rangle$ with momentum $k$ where the
 DW centers have distance $d=1$ (see Fig.~\ref{fig:dws_p}(b))
\begin{equation}\begin{split}
|\psi_{d=1,k}\rangle &= {c}^\dagger_{k, \uparrow} |\phi_\textrm{WL}\rangle =
 \frac{1}{\sqrt{L}} \sum_{r=1}^{L} \textrm{e} ^{i kr}{c}^\dagger_{r,
 \uparrow} |\phi_\textrm{WL}\rangle = \\
& = \frac{1}{\sqrt{2L}} (\textrm{e} ^{i k}|\uparrow \uparrow 0 \uparrow 0
 \uparrow \dots \rangle - \textrm{e} ^{2 i kr} |\uparrow  \uparrow \uparrow 0
 \uparrow 0 \dots \rangle \\
 &\quad\quad\quad - \textrm{e} ^{3 i k}|0 \uparrow \uparrow \uparrow 0
 \uparrow \dots \rangle + \dots )\;,
\end{split}\end{equation}
where the $-$ signs are due to a Fermi sign obtained by moving the creator
through an existing electron.
Once the two DWs have been created at distance $d=1$, they can move 
via NN electron hopping $t_1$:
\begin{equation}\begin{split}\label{eq:hopp_dws}
T_1|\psi_{d=1,k}\rangle &= t_1 \sum_{r=1}^{L} ({c}^\dagger_{r, \uparrow} 
{c}^{\phantom{\dagger}}_{r+1, \uparrow} + \textrm{h.c.})\ |\psi_{d=1,k}\rangle = \\
& = \frac{t_1}{\sqrt{2L}} \left(\textrm{e} ^{i k}(|\uparrow 0 \uparrow \uparrow 0
 \uparrow \dots \rangle + |\uparrow \uparrow 0 \uparrow 
 \dots \uparrow \uparrow  0 \rangle) \right.\\
 &\quad - \textrm{e} ^{2 i kr} (|\uparrow  \uparrow 0 \uparrow
 \uparrow 0 \dots \rangle + |0  \uparrow \uparrow
 0 \uparrow 0 \dots \uparrow \uparrow \rangle) \\
 &\quad- \textrm{e} ^{3 i k}(|\uparrow 0 \uparrow \uparrow 0
 \uparrow \dots \rangle + |0 \uparrow \uparrow 0 \uparrow
 \uparrow \dots \rangle) \\&\left.\quad + \dots \right) =\\
 & = 2 t_1\sin k|\psi_{d=3,k}\rangle \;,
\end{split}\end{equation}
where $|\psi_{d=3,k}\rangle$ denotes the state with momentum $k$ and distance
$d=3$ between the DWs (a complex phase has been absorbed into its
definition). Repeating step (\ref{eq:hopp_dws}), $d$ can grow further, which each
$t_1$ process increasing (or reducing) $d$ in steps of two, see
Fig.~\ref{fig:dws_p}. (Additionally, $t_1$ can introduce more DWs, as in
Fig.~\ref{fig:dws_pm}(b), but such processes cost energy and will be neglected in this
low-energy analysis.) 

We now arrive at the effective DW Hamiltonian $\mathcal{H}_\textrm{DW}$ in
terms of $d$ and $k$. 
\begin{equation}\label{eq:ham_dw} 
\mathcal{H}_\textrm{DW} = \left( \begin{array}{cccccc}
     E_1       & \tilde t_1 (k) & 0              & \dots          &       &\\
\tilde t_1 (k) &      E_3       & \tilde t_1 (k) & 0              & \dots & \\
0              & \tilde t_1 (k) & E_5            & \tilde t_1 (k) & 0  & \dots\\\
\vdots         &                &                &\ddots &&
\end{array}\right)\;,
\end{equation}
where the diagonal contains the
potential energy $E_d$ and momentum $k$ enters the effective hopping $\tilde
t_1 (k) = 2 t_1\sin k$.

\begin{figure}
\centering
\subfigure{\includegraphics[width=0.4\textwidth]
    {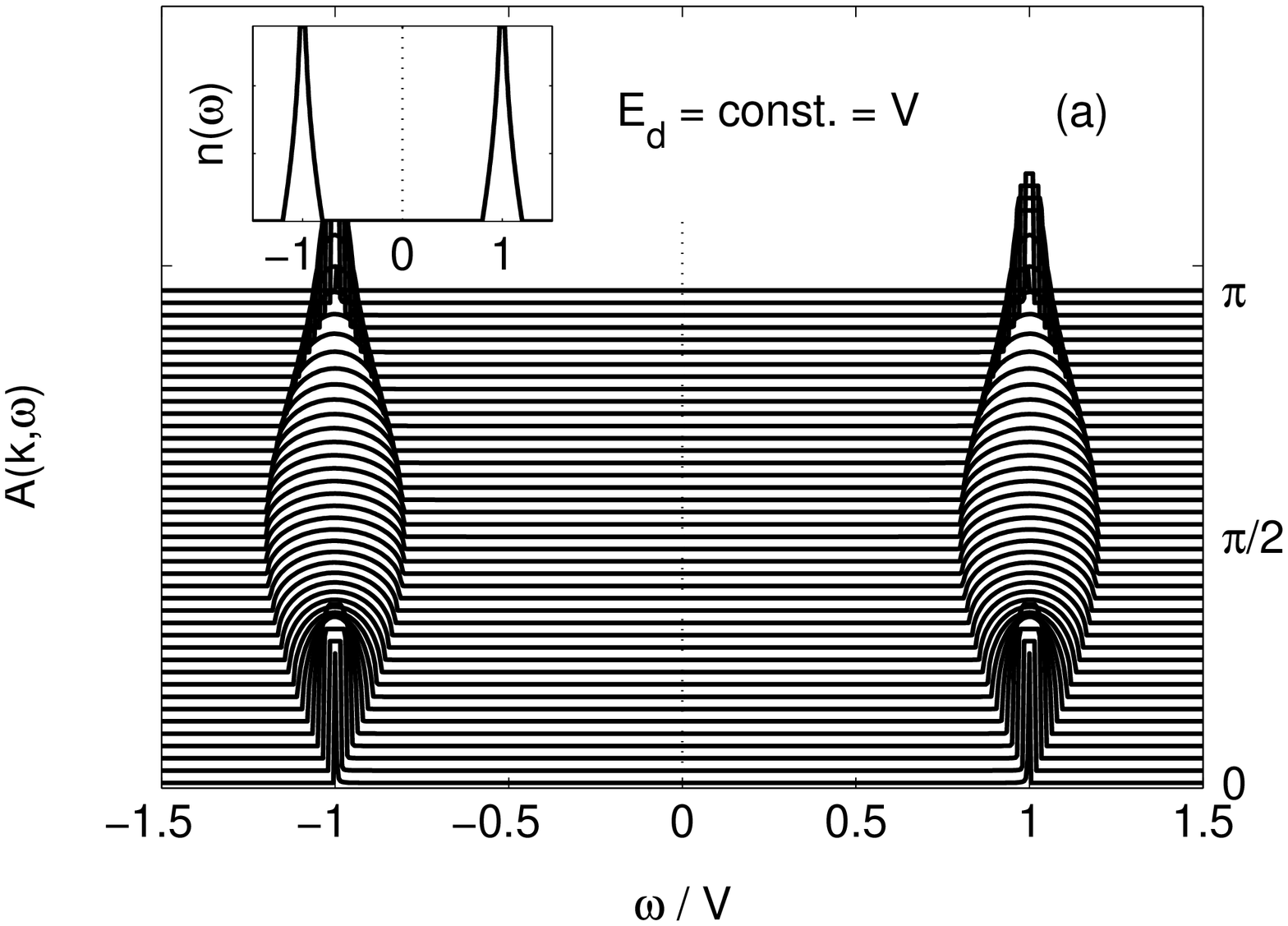}\label{fig:spec_DW_lmax1}}\\[-3.5em]
\subfigure{\includegraphics[width=0.4\textwidth]
    {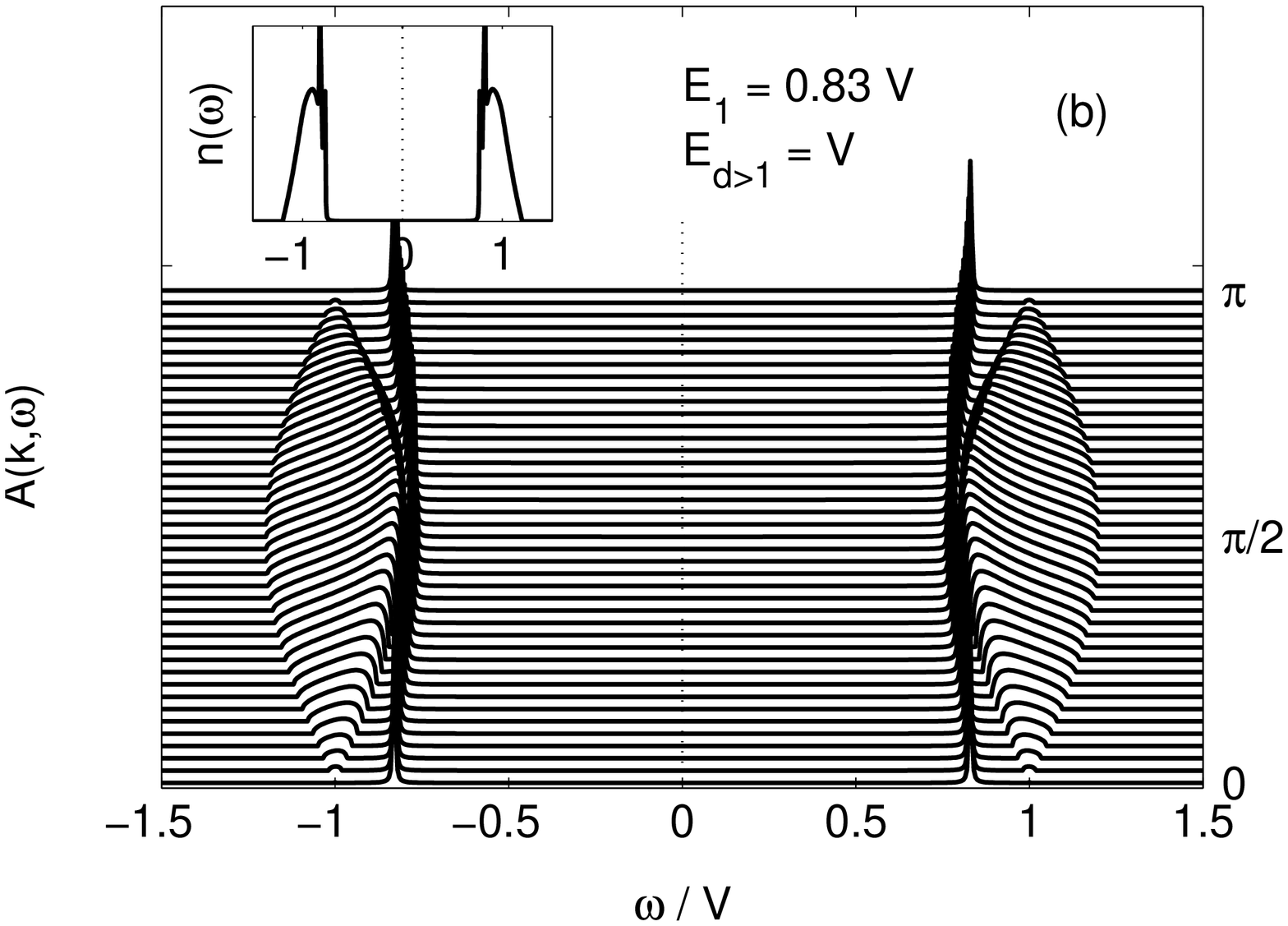}\label{fig:spec_DW_lmax3}}\\[-3.5em]
\subfigure{\includegraphics[width=0.4\textwidth]
    {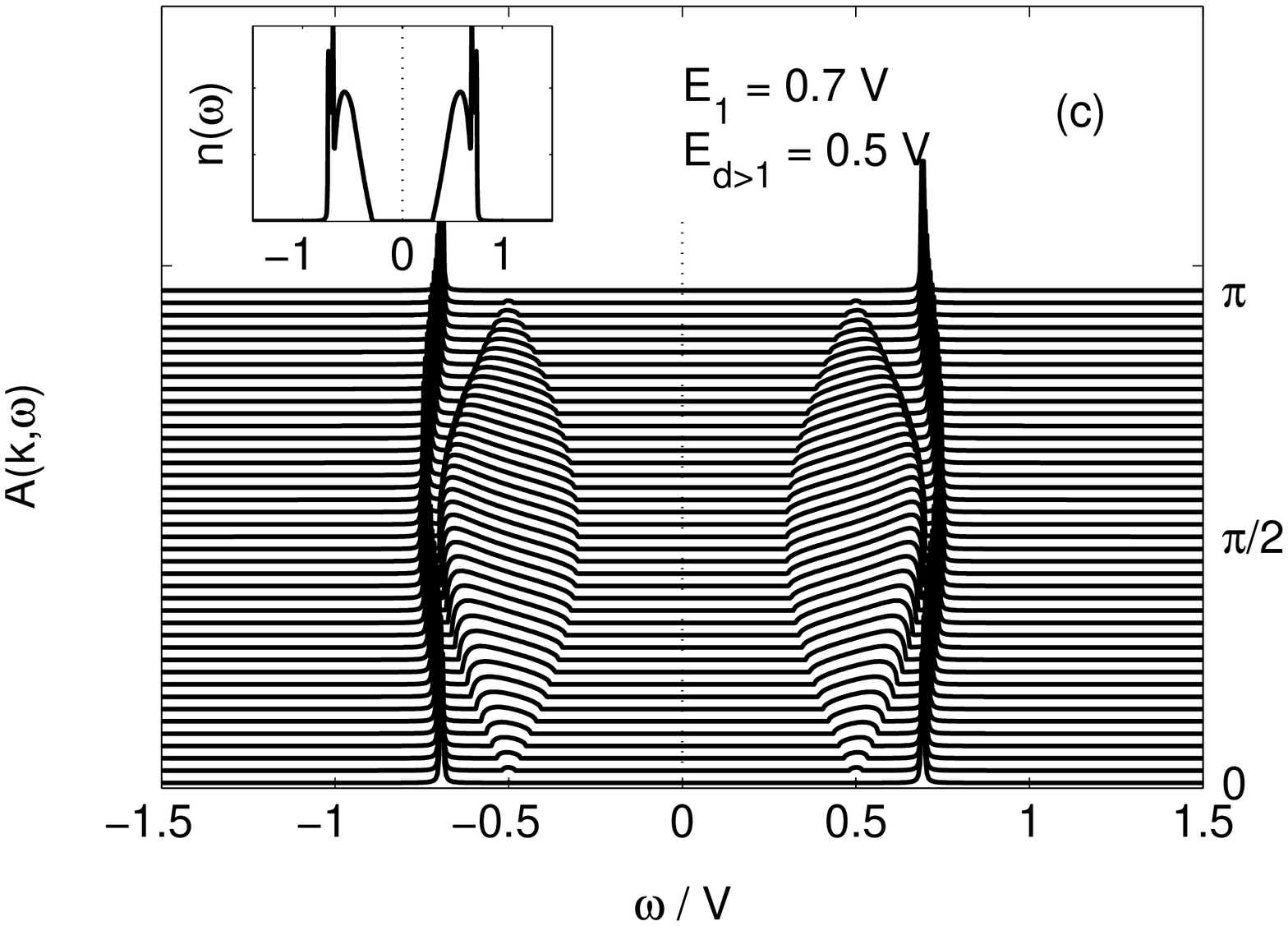}\label{fig:spec_DW_lr}}\\[-1.5em]
\caption{Spectral density obtained from the effective two-DW Hamiltonian
  (\ref{eq:ham_dw}) for $t_1 = -0.05 V, t_2 = 0$. (a) For non-interacting DWs with $E_d =
  \textrm{const}$, see Fig.~\ref{fig:spec_spinless_lmax1} for the spectral density for
  the corresponding case $l_\textrm{max}=1$ in the electron Hamiltonian
  (\ref{eq:hamiltonian}). (b) Attractive DW interaction with $E_1 = 0.83 V,
  E_{d\geq 3} = V$, resulting from $l_\textrm{max}=3$ in
  (\ref{eq:hamiltonian}), compare to  Fig.~\ref{fig:spec_spinless_lmax3}. (c)
  Repulsive DW interaction with $E_1 =0.7 V,  E_{d\geq 3} = 0.5 V$ as an
  approximation to the DW potential of the long-range electron Hamiltonian with
  $l_\textrm{max} = L/2$, the pertaining spectral density is seen in
  Fig.~\ref{fig:spec_spinless_coulomb}. The $\delta$ peaks 
 of  bound states have been slightly
  broadened and their height has been cut for better visibility
  of the 2 DW-continua.\label{fig:spec_DW}}
\end{figure}

We are interested in the spectral density (\ref{eq:ak}), which can be written as
\begin{equation}\begin{split}
A( k, \omega)&  =  -\frac{1}{\pi} \Im \langle \phi_0 | {c}^{\phantom{\dagger}}_{ k, \uparrow}
\frac{1}{\omega - (H-E_0) + i0^+}{c}^\dagger_{ k, \uparrow}| \phi_0\rangle \\
& \approx -\frac{1}{\pi} \Im \langle \psi_{1,k} | 
\frac{1}{\omega - \mathcal{H}_\textrm{DW} + i0^+}|\psi_{1,k}\rangle\;,
\end{split}\end{equation}
i.e., we need the $(1,1)$ element of the inverse of the matrix $\omega -
\mathcal{H}_\textrm{DW}$, which can be written as a continued fraction
\begin{equation}
\label{eq:chain_fract}
(\omega -\mathcal{H}_\textrm{DW})^{-1}_{1,1} = \frac{1}{\omega - E_1 -
  \frac{\tilde t_1(k)^2}{\omega - E_3 - \frac{\tilde t_1(k)^2}{\omega - E_5 - \dots} } }\,.
\end{equation}
This fraction can be easily evaluated for the distance independent DW
potential $E_d=V$ that is found, if only NN Coulomb repulsion is kept
in the electronic Hamiltonian (\ref{eq:hamiltonian}), see
Fig.~\ref{fig:pot_dw}. We then obtain
\begin{equation}
\label{eq:chain_fract_solved}
A_1=(\omega -\mathcal{H}_\textrm{DW})^{-1}_{1,1} = \frac{\omega - \tilde E}{ 2\tilde
  t_1(k)^2} \pm \sqrt{\left(\frac{\omega - \tilde E}{ 2\tilde
  t_1(k)^2}\right)^2-\frac{1}{\tilde t_1(k)^2}}\,,
\end{equation}
where $\tilde E$ is denotes the distance-independent DW potential, here $\tilde E = V$.
The resulting spectral density has only a regular part and no singularities
(except for $k=0,\pi$, where the hopping $\tilde t_1(k)$ vanishes), and is shown in
Fig.~\ref{fig:spec_DW_lmax1}. The incoherent two-DW continuum at energies $V-
4t_1\sin k < \omega < V+ 4t_1\sin k $ results from the independent (since $E_d
= \tilde E = \textrm{const}$) movement of the two DWs, and corresponds to the
spectral density depicted in Fig.~\ref{fig:spec_spinless_lmax1}, which was
calculated for the full fermionic model (\ref{eq:hamiltonian}) with only NN
repulsion ($l_\textrm{max}=1$).

Next, we move to the case $l_\textrm{max}=3$, which leads to a short range
attraction between the DWs, see Fig.~\ref{fig:pot_dw}. Since the potential is
constant for $d\geq 3$, we can use the result obtained above for $E_d =
\textrm{const}$ in order to arrive at
\begin{equation}
\label{eq:chain_fract_solved_3}
A_3=\frac {1}{\omega - E_1 - \tilde t_1(k)^2 A_1}\,,
\end{equation}
which is shown in Fig.~\ref{fig:spec_DW_lmax3}. In addition to the continuum,
we now find a pole where the denominator of
(\ref{eq:chain_fract_solved_3}) vanishes. It results from the DW attraction
and corresponds to a bound quasiparticle with dispersion
\begin{equation}\label{eq:disp}
\epsilon( k) = E_1 + \frac{\tilde t_1(k)^2}{E_1 - \tilde E}= E_1 + \frac{(2t_1\sin k )^2}{E_1 - \tilde E}\;, 
\end{equation}
where $\tilde E = V$ is the potential at distances $d\geq 3$ and $E_1 \sim0.83
V$ the potential at $d=1$. For long-range electron-electron repulsion, we
simplify the resulting DW potential to $E_1 = 0.7V, \tilde E = 0.5V$ and can
then apply (\ref{eq:chain_fract_solved_3}) and (\ref{eq:disp}). However, the
quasiparticle is now an \emph{antibound} state \emph{above} the incoherent
continuum, see Fig~\ref{fig:spec_DW_lr}. 
This highly unusual situation is also
observed in the spectra of the full model ( see Fig.~\ref{fig:spec_spinless_coulomb})
obtained by exact diagonalization. 
As we
have seen here, it results from the DW repulsion and is thus a  direct
consequence of the long-range nature of the Coulomb repulsion absent from
short-range models. 

The only feature of the spectra for the full model (\ref{eq:hamiltonian}) that
is not reproduced in the simplified DW analysis is the transfer of spectral
weight seen in Fig.~\ref{fig:spec_spinless} with more weight in
photoemission (PES) at $k=0$ and more in inverse photoemission (IPES) at
$k=\pi$. This transfer is caused by fluctuations around 
perfect charge order in the ground state induced by $t_1$, see
Fig.~\ref{fig:dws_pm}, and will be discussed elsewhere. 

\begin{figure}
\centering
\subfigure{\includegraphics[width=0.4\textwidth]
    {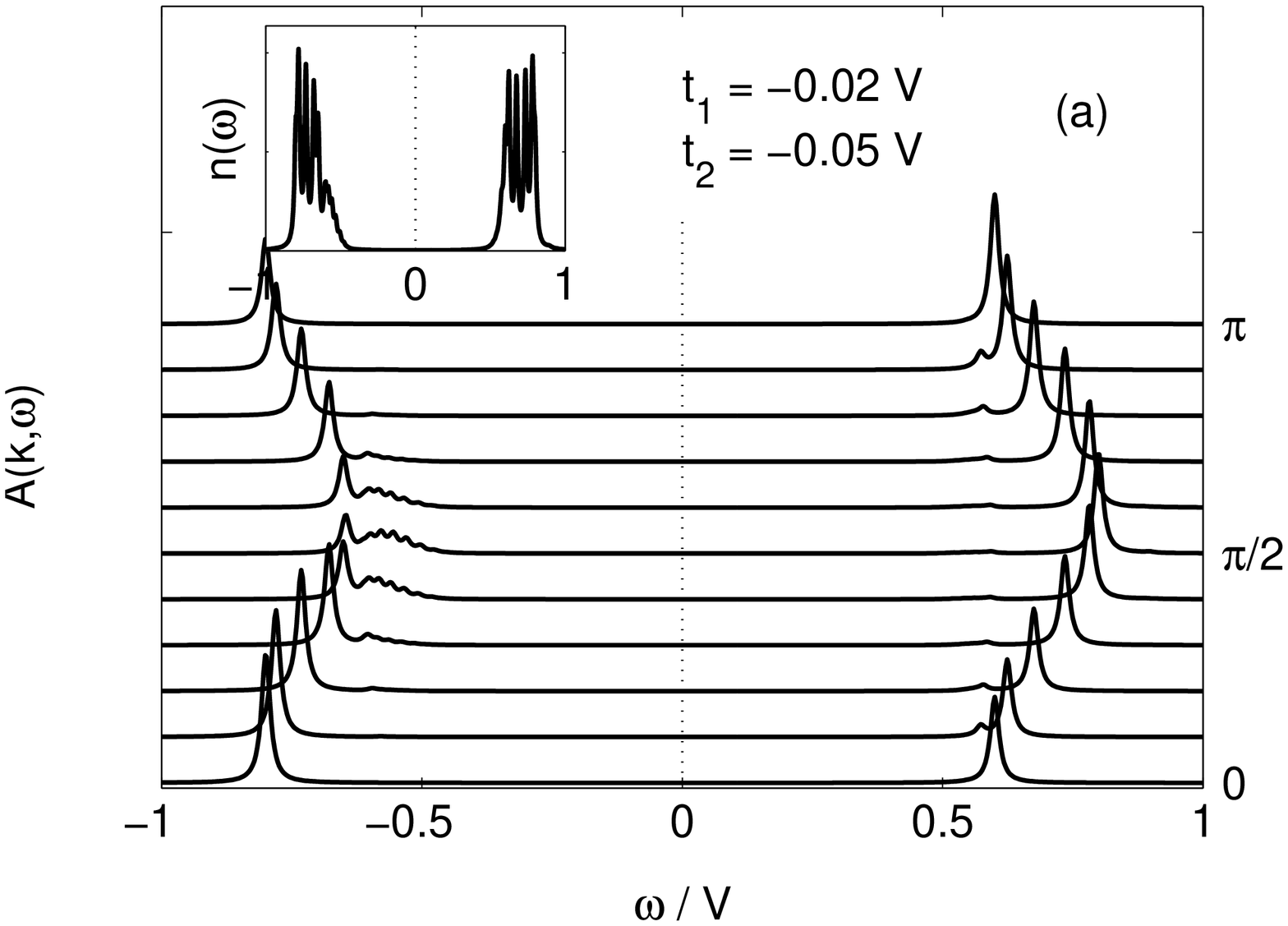}\label{fig:spec_t1_t2_spinless}}\\
\subfigure{\includegraphics[width=0.4\textwidth]
    {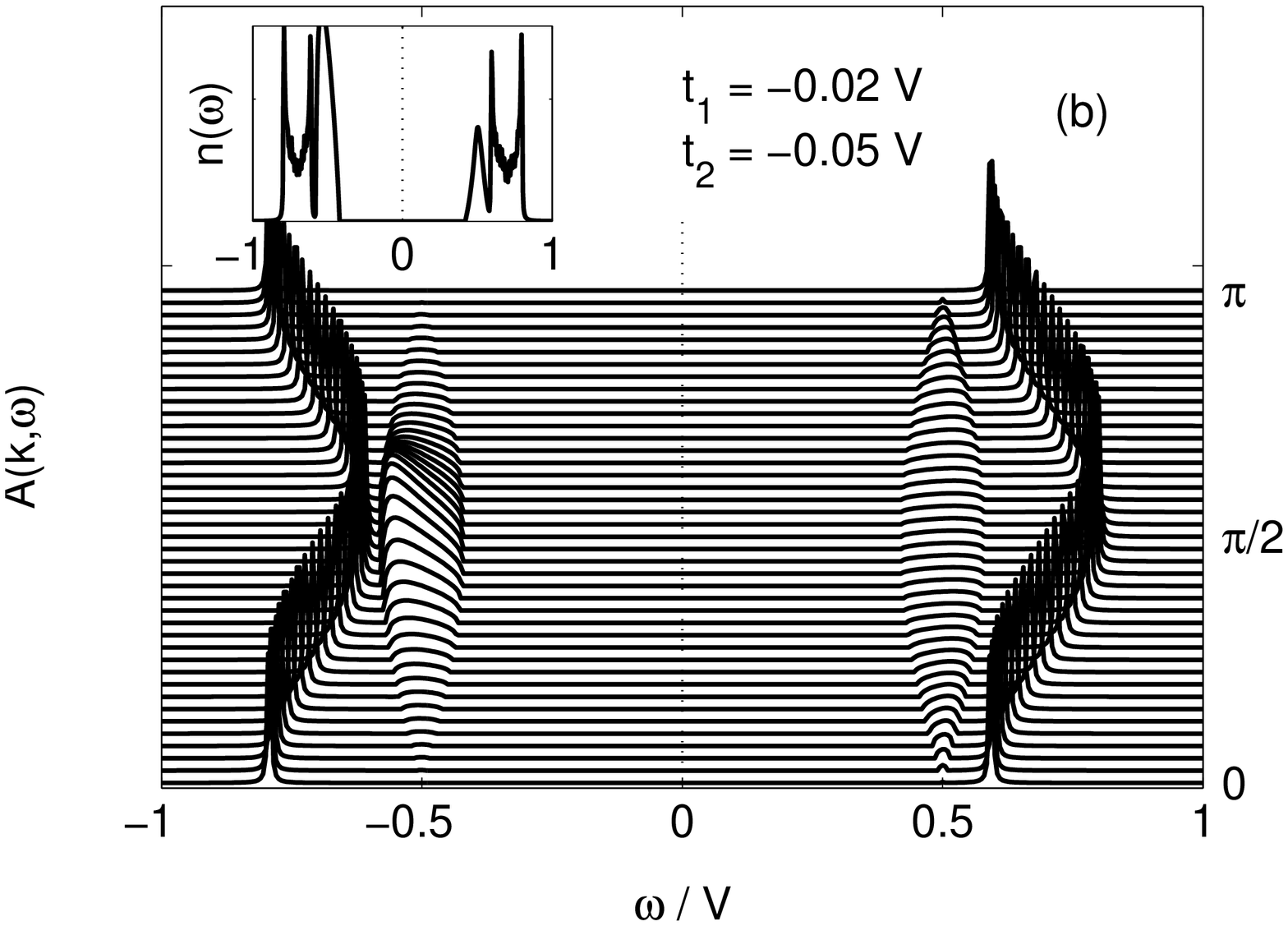}\label{fig:spec_t1_t2_DW}}\\
\caption{Spectral density $A(k,\omega)$ for $t_1 = -0.02 V, t_2  
  = -0.05V$. (a) For the spinless-fermion Hamiltonian (\ref{eq:hamiltonian}) with long-range interaction
  $l_\textrm{max} = L/2$ obtained by Lanczos diagonalization via (\ref{eq:ak}). (b) For the effective
  DW-Hamiltonian (\ref{eq:ham_dw}) obtained by (\ref{eq:chain_fract_solved_3}) with $E_1 = 0.7V
  + 2t_2\cos 2k , \tilde E = 0.5V$. Insets show the density of states.\label{fig:spec_t1_t2}}
\end{figure}

Motivated by experimental data indicating that NNN hopping $t_2$ is larger
than $t_1$ in Na$_3$Cu$_2$O$_4$,~\cite{Horsch05} we now turn to the spectral
density for $t_2 \neq 0$. Via NNN hopping, an electron inserted into an empty
site of the WL can hop over the occupied sites and move along the empty ones
like a free fermion, and the same holds for a hole inserted into an occupied
site. For $t_1=0$, we therefore obtain an independent-electron-like dispersion
with $\epsilon(k) = V \ln 2 + 2t_2\cos 2k$ (not shown). The dispersion is the
same in PES and IPES, which results in an indirect band gap with the highest
occupied state at $k=\pi/2$ and the lowest unoccupied ones at $k = 0, \pi$. Finite NN hopping
$t_1\neq 0$ can be taken into account just like in the $t_2 = 0$ case
discussed above by setting $E_1 = V \ln 2 + 2t_2\cos 2k$ in (\ref{eq:chain_fract_solved_3}). The
resulting spectral density of the effective DW model is shown in
Fig.~\ref{fig:spec_t1_t2_DW} and the corresponding results for the fermion
Hamiltonian (\ref{eq:hamiltonian}) in
Fig.~\ref{fig:spec_t1_t2_spinless}. Both for PES and for IPES, we see 
the 2-DW continuum as well as  a quasiparticle 
dispersion $\sim \cos 2k$. But we observe a certain particle-hole asymmetry: For
IPES, the continuum has small weight and the width of the dispersion is
$4t_2$ as for a free electron. For PES, the continuum is stronger and almost mixes
with the quasiparticle at momenta $k\sim \pi/2$, which somewhat reduces the
band width for the quasiparticle. We will see in the next section that the
particle-hole asymmetry is strongly enhanced for electrons with spin.

\section{Dynamics of electrons with spin}

\label{sec:spin}
\begin{figure}[t]
  \centering
  \psfrag{a}{\hspace*{-1em}(a)}
  \psfrag{b}{\hspace*{-1em}(b)}
  \psfrag{c}{\hspace*{-1em}(c)}
  \psfrag{d}{\hspace*{-1em}(d)}
  \psfrag{t}{$t_1$}
  \includegraphics[width=0.4\textwidth]{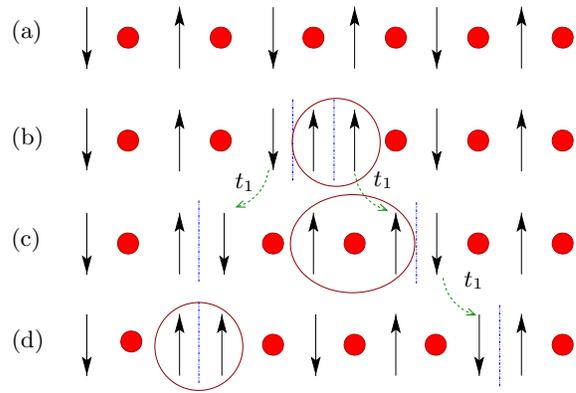}\\
\caption{(Color online) An electron added to the perfect WL, as
  Fig.~\ref{fig:dws_p}, but for electrons with spin. In addition to the two
  DWs, an electron now creates a spinon (denoted by the unfilled
  circle/ellipse). As in the spinless case, the DWs can move via $t_1$, and
  the spinon can move via spin-flip processes. \label{fig:dw_spinon}}
\end{figure}

\begin{figure}
  \centering
  \vspace*{-0.5em}
  \includegraphics[width=0.4\textwidth]{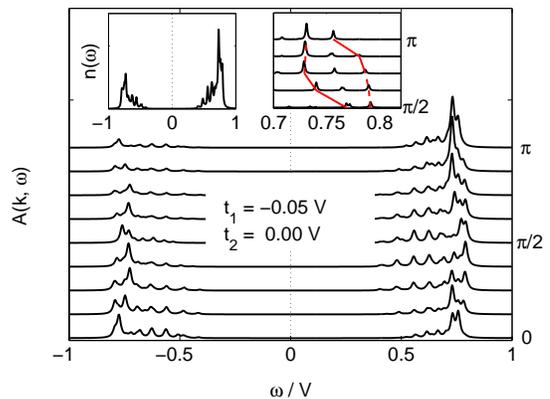}\\[-0.5em]
  \caption{(Color online) Spectral density $A(k,\omega)$ with spin for $t_1 =-0.05 V,
    t_2 = 0$, $U = 4V$, $L = 16$. The right-hand inset shows a blow-up of IPES with
    a higher energy-resolution, the dashed and solid lines indicate 
    spinon and `anti-holon' branches (guides to the eye). The inset on the left shows the density of 
    states.\label{fig:spec_spin_t1} }
\end{figure}

\begin{figure}
  \centering
  \vspace*{-0.5em}
  \subfigure[]{\includegraphics[width=0.23\textwidth]
    {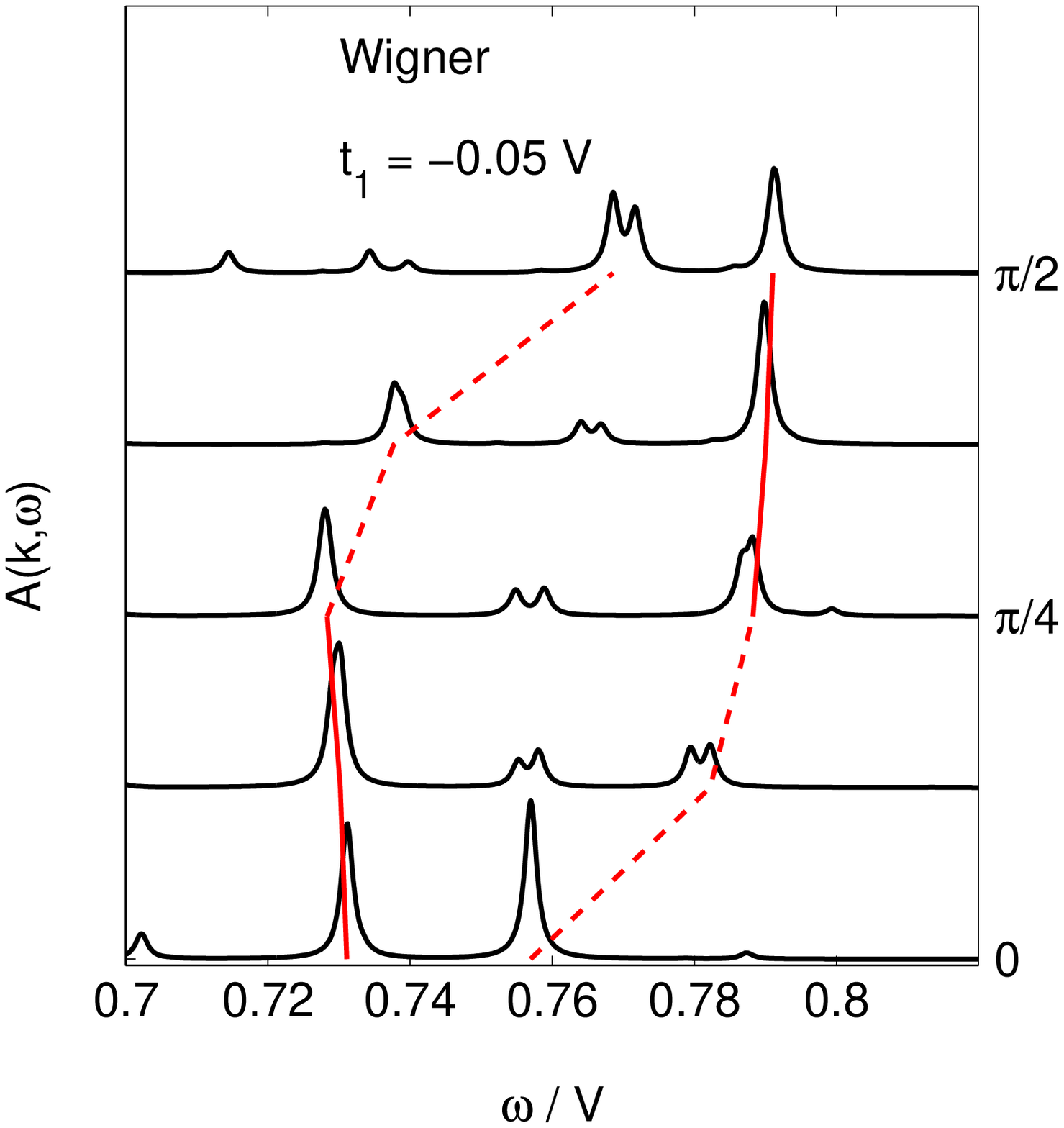}\label{fig:wigner_blowup}}
  \subfigure[]{\includegraphics[width=0.23\textwidth]
    {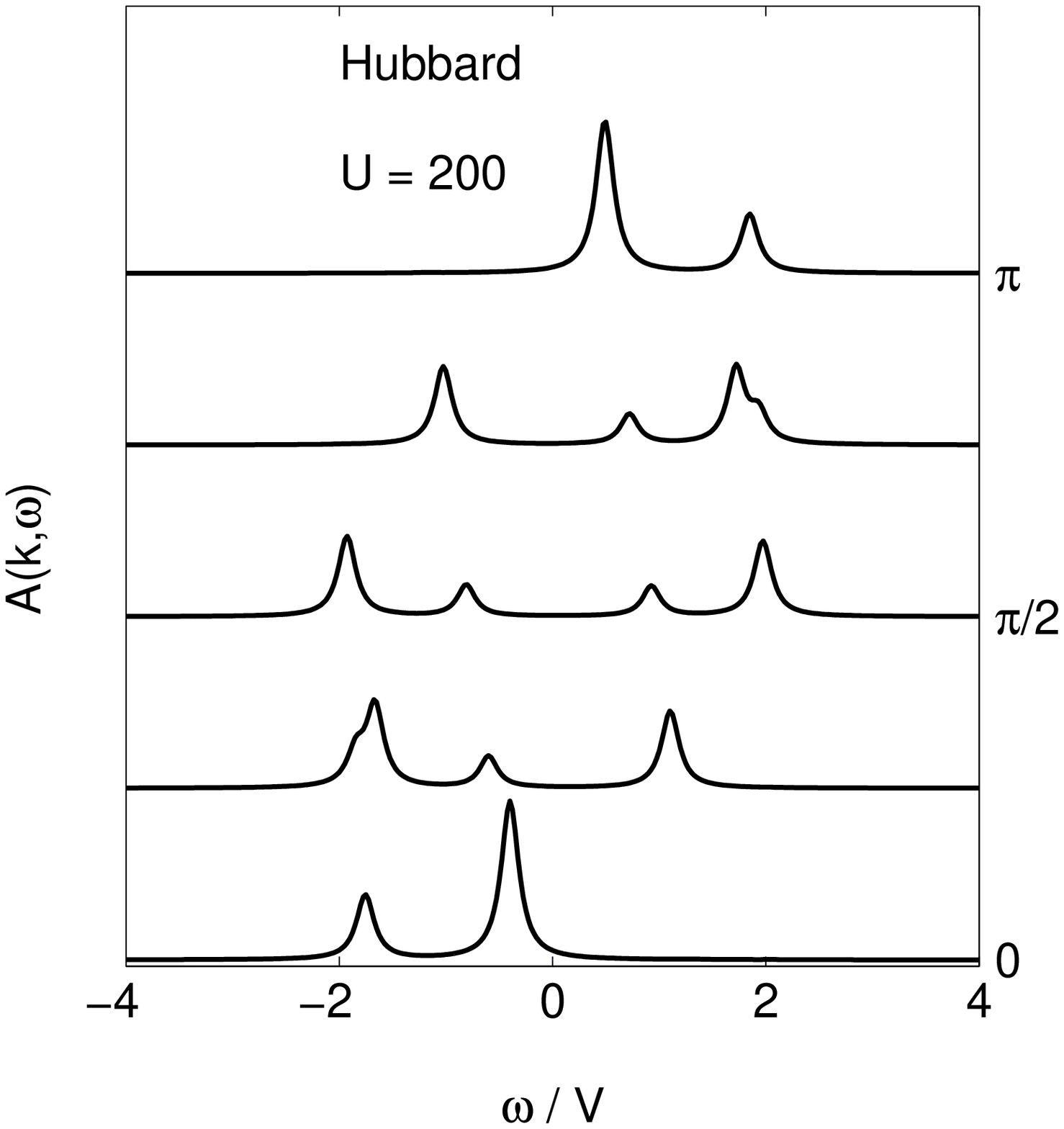}\label{fig:hubbard}}\\[-0.5em]
  \caption{Comparison of the spectral density for Wigner and Hubbard
    models. (a) Blow-up with higher energy resolution of IPES for the Wigner
    model ($L=16$ sites) as in Fig.~\ref{fig:spec_spin_t1}. 
   The dashed and solid lines are
    guides to the eye indicating spinon and `anti-holon' branches. (b) IPES
    for the half-filled 1D Hubbard model 
    ($L=8$) with NN topping $t=1$ and $U=200t$.
    \label{fig:comp_wigner_hubb}}
\end{figure}

Having understood the dynamics of the spinless system, i.e., of the
charge sector, we now include the spin. An inserted electron or hole then produces
a spinon in addition to the two DWs, see the cartoon
in Fig.~\ref{fig:dw_spinon} illustrating the possible processes in the WL with
spin. Again, the dynamics of the
charge sector are determined by DW hopping via $t_1$. The spinon can move by
spin-flip processes $\sim J_2 = t_2^2/U$ or $\sim J_1 = t_1^2/U$, a 
magnetic scale much smaller than the hopping scale. In this section, we will
discuss how the spectral density indicates spin-charge separation similar as in
the half-filled 1D Hubbard model.

\begin{figure}
\centerline{
\includegraphics[width=0.48\textwidth]{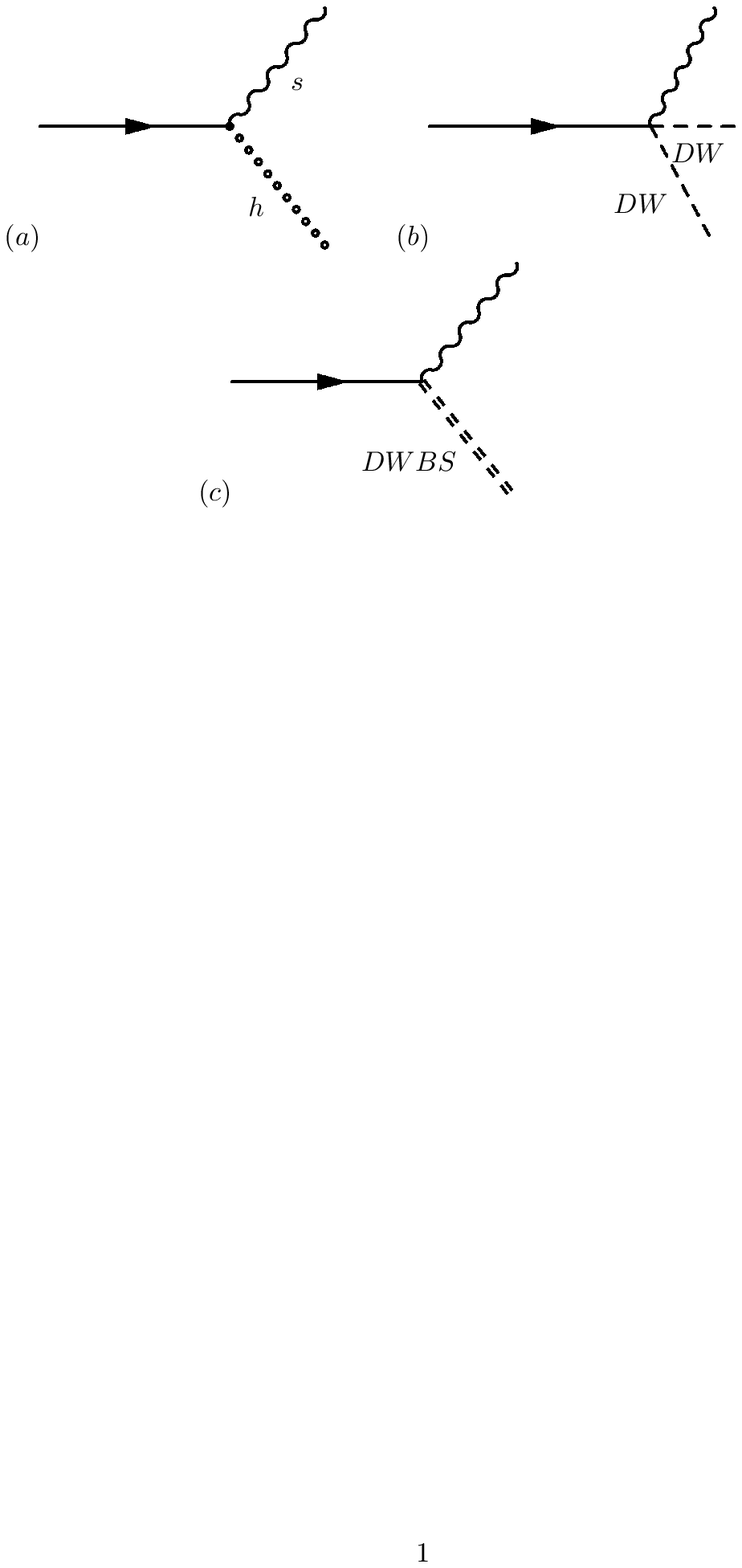}}
\caption{(a) Charge-spin separation in 1D Hubbard model:
Electron (solid line) decays into holon and spinon;
(b) Wigner lattice at quarter-filling: Decay of electron
into 2 DWs and a spinon;
(c) Long-range Coulomb-repulsion of DWs generates a  DW (anti) bound state (DWBS).
This scattering state behaves like a holon. \label{fig:vertex}}
\end{figure}

At first, we will  focus on NN hopping and compare the spectral density
for $t_1 = -0.05 V, t_2=0$  with spin (Fig.~\ref{fig:spec_spin_t1}) to the one
without spin (Fig.~\ref{fig:spec_spinless_coulomb}). The 2-DW continuum is
broader with spin, e.g. at $k=0$ and$k=\pi$, where the width of the continuum 
shrinks to zero in the spinless model.
A more fundamental change is apparent in the antibound quasi-particle of
the spinless case: It has evolved into a narrow structure with high spectral
weight comprised of several peaks. Figure~\ref{fig:wigner_blowup} shows a
blow-up with higher energy resolution of this structure (in IPES) for $k=0$ to
$k=\pi/2$. When we compare this blow-up to IPES for the usual
half-filled 1D Hubbard model ($t=1, U=200t$) shown in Fig.~\ref{fig:hubbard},
we observe remarkably similar structures. It is well known that an electron in the
Hubbard model decays into a spinon and an anti-holon, see
Fig.~\ref{fig:vertex}(a), and that IPES is given by a convolution of spinon
and anti-holon branches. While this may not be so obvious on the
present short 8-site ring, the anti-holon with width $\sim t$ and the spinon
$\sim J=4t^2/U \ll t$ are clearly visible in the thermodynamic
limit.\cite{Aichhorn03} The strong similarities seen in the spectral densities
Fig.~\ref{fig:comp_wigner_hubb} for the Wigner and the Hubbard models let us
conclude that the Wigner model likewise shows a convolution of spinon and
anti-holon bands; the role of the anti-holon is now taken by the antibound
2-DW state that is the elementary collective excitation of the charge sector, see
Fig.~\ref{fig:vertex}(c). Since the antibound 2-DW state already has a
periodicity of $\pi$, see (\ref{eq:disp}), the momentum range $k=0$ to
$k=\pi/2$ corresponds to $k=0$ to $k=\pi$ in the Hubbard model with NN
hopping, where the one-particle dispersion has periodicity $2\pi$. The width
of the convolution is determined by that of the antibound 2-DW state,
hence $\sim \frac{(2t_1\sin k )^2}{E_1 - \tilde E}$, see (\ref{eq:disp}); the
spinon has almost no dispersion, because $\sim J_1 \ll t_1$. The small
differences between the Figs.~\ref{fig:wigner_blowup} and~\ref{fig:hubbard}
are due to weak interactions with the 2-DW continuum in the former case.

\begin{figure}[h]
  \centering
  \vspace*{-0.5em}
  \subfigure{\includegraphics[width=0.4\textwidth]
    {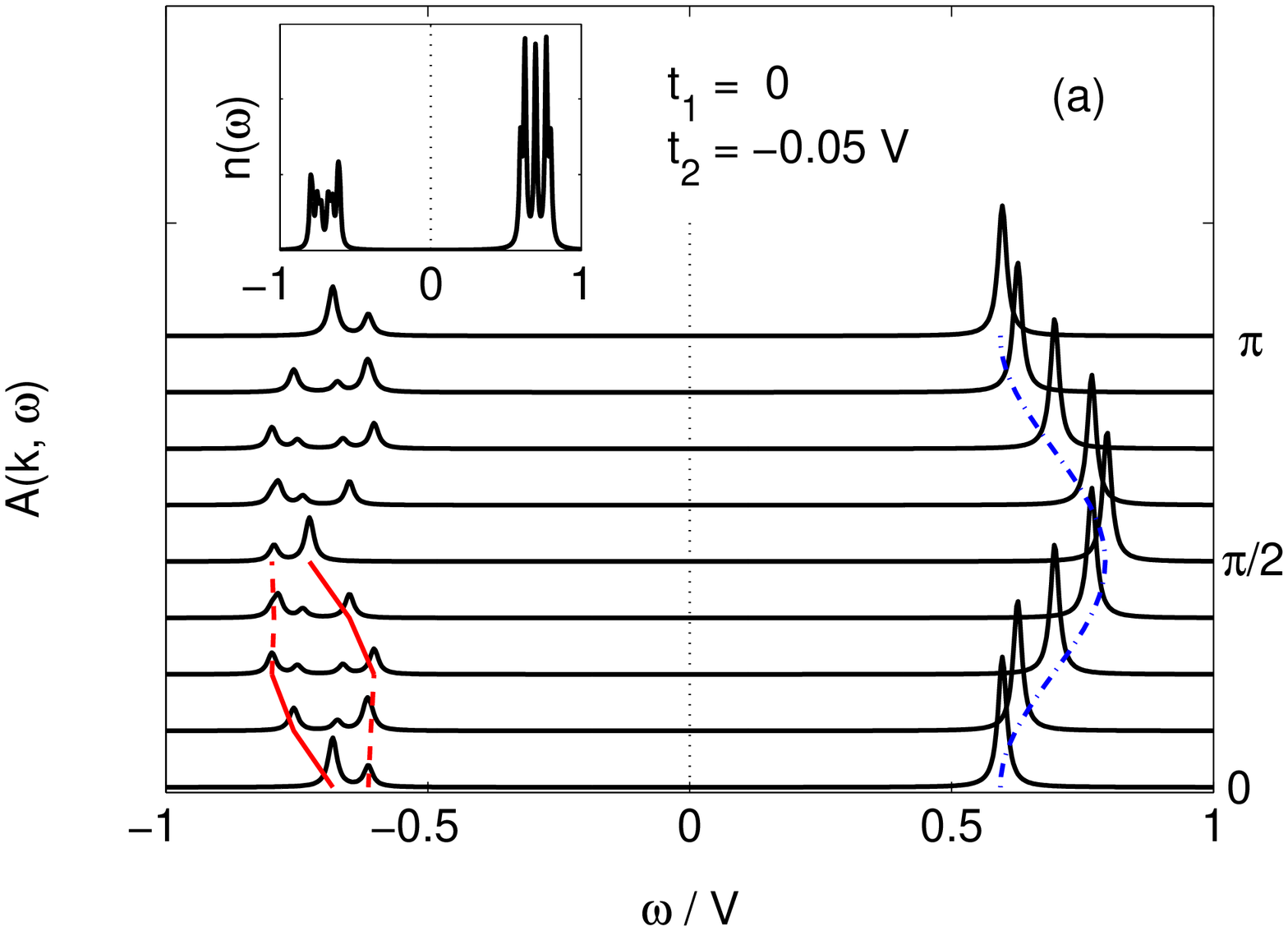}\label{fig:spec_spin_t2}}\\[-3.5em]
  \subfigure{\includegraphics[width=0.4\textwidth]
    {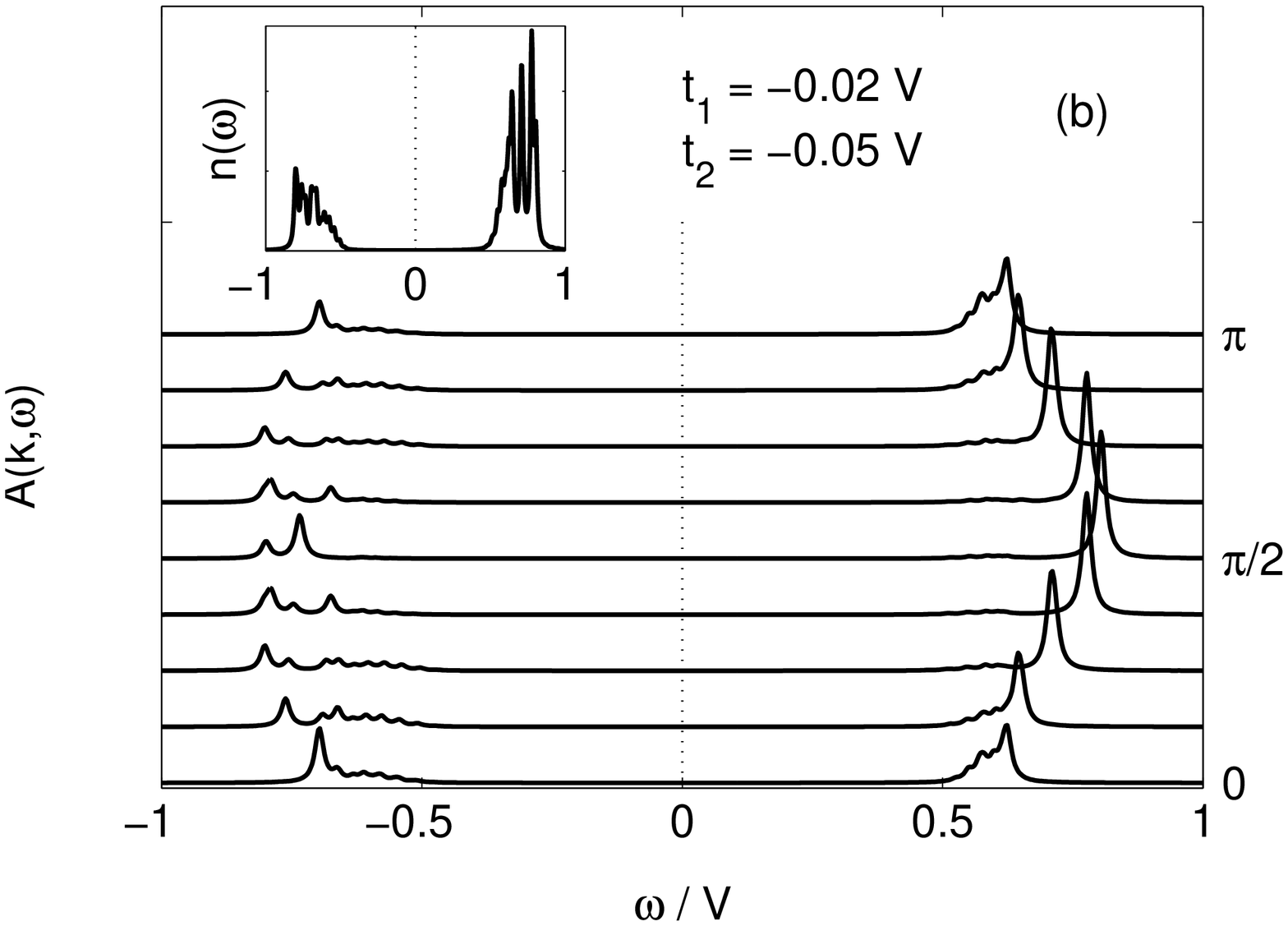}\label{fig:spec_spin_t1_t2}}\\[-1.5em]
  \caption{(Color online) Spectral density $A(k,\omega)$ with spin for (a)
    $t_1 = 0, t_2 = -0.05V$, (b) $t_1 = -0.02 V, t_2 = -0.05V$, 
    $U = 4V$, $L = 16$. Insets on the left show the density of 
    states.  In (a), spinon and holon branches in PES are indicated by
    dashed and solid lines (guides to the eye), the dash-dotted line in 
    IPES gives the one-particle dispersion $E(k) = V \ln 2 +
    2t_2\cos 2k$. \label{fig:spec_spin_2}} 
\end{figure}

We now turn to the spectral density 
for $t_2 \neq 0$, at first choosing $t_1 =0$ for simplicity. The ground state is then
given by the perfectly charge ordered WL, an electron added in IPES goes into an empty
site and can move freely on the empty sublattice, without any interaction with the
occupied sublattice (as $t_1 =0$). Consequently, IPES shows a 
one-particle-like tight-binding band with dispersion $E(k) = V \ln 2 +
2t_2\cos 2k$ just like in the spinless model, see $\omega>0$ in Fig.~\ref{fig:spec_spin_t2}. The situation
for a hole is, however, fundamentally different: The hole goes into the occupied sublattice,
which is a half-filled \emph{Hubbard} chain, and it therefore separates into spin and
charge. The resulting spinon and holon branches in PES can be seen for
$\omega < 0$ in Fig.~\ref{fig:spec_spin_t2}. (Again, the spectrum was found
to agree to the one for the Hubbard model on eight sites. In this case even
without \emph{any} deviation, because there is no 2-DW continuum.) In this
case of extreme particle-hole asymmetry, one
and the same observable, the spectral density, shows both pure
one-particle behavior (in the particle sector) and strongly correlated
behavior (in the hole sector). 
If both $t_2$ and $t_1$ are active, the inserted particle (hole) can interact with the occupied
(empty) sub-lattice, see the spectral density in
Fig.~\ref{fig:spec_spin_t1_t2}. $t_1$-hopping processes now induce an
additional incoherent 2-DW continuum, and we aquire incoherent weight
in IPES. However, the strong particle-hole asymmetry persists: As in the
spinless case shown in Fig.~\ref{fig:spec_t1_t2}, we  
see that the quasiparticle in IPES remains largely intact and that -
surprisingly - the peaks furthest from the Fermi energy remain sharpest.

\section{Summary and Conclusions}
\label{sec:conclusions}

To conclude, we have investigated the dynamics of a quarter-filled
Hubbard-Wigner model. Apart from being the most transparent situation on a
Wigner lattice, quarter-filling is also appropriate for the edge-sharing
chain compound Na$_3$Cu$_2$O$_4$. We find that electrons decay into a spinon
and two domain walls as sketched in Fig.~\ref{fig:vertex}(b). 

In the absence of the spinon, i.e., for the spinless model, we have
investigated the effects of a truncated Coulomb interaction vs. fully
long-range interaction in Sec.~\ref{sec:spinless}. By comparing dynamic
observables of the Wigner model (\ref{eq:hamiltonian}) to those
of an effective DW Hamiltonian, we have shown that the DWs and their
interaction clearly manifest themselves in observables and are thus accessible
to experiment. In the case of truly long-range electron-electron interaction,
the formal fractional charges can be given a direct physical meaning that is
consistent for both one-particle spectra $A(k, \omega)$ and two-particle
dynamics $N(q,\omega)$. In models with truncated interaction, we still find
formal fractional charges, but their interaction does no longer directly
correspond to their formal charge. This difference has a strong impact on
$A(k\omega)$: The DW repulsion due to long-range electron-electron interaction
leads to a highly unusual \emph{anti-}bound quasiparticle vs. a more
conventional bound quasi-particle observed for truncated electron-electron
interaction. 

We have finally analyzed the Hubbard-Wigner model for electrons with spin, where we
find the signatures of both charge fractionalization and spin-charge
separation. Despite its composite nature, the antibound
quasiparticle undergoes spin-charge separation reminiscent of an electron in
the 1D Hubbard model, see Fig.~\ref{fig:vertex}(c). Experimentally, $t_2 >
t_1$ could be more relevant,~\cite{Horsch05} and this case shows striking
particle-hole asymmetry: For $t_1=0$, an added electron behaves like an
independent particle while a hole shows strongly correlated behavior and
decomposes into spinon and holon. Even for finite $t_1$, this behavior
persists to some extend.

We thank A.~M.~Ole\'s, J.~Unterhinninghofen and O.~Gunnarsson
for careful reading of the manuscript and for helpful suggestions.


\begin{thebibliography}{10}
\bibitem{note2}
\textnormal{As we treat rings with length $L$, our interaction is in fact
  $V_l=\max(V/l,V/(L-l))$, with $l_{\textrm{max}} = L/2$}.
\end{thebibliography}

\end{document}